# High-order Photonic Cavity Modes Enabled 3D Structural Colors

*Hailong Liu, Hongtao Wang, Hao Wang, Jie Deng, Qifeng Ruan, Wang Zhang, Omar A. M. Abdelraouf, Noman Soo Seng Ang, Zhaogang Dong, Joel K. W. Yang\*, and Hong Liu\**

H.L. Liu, J. Deng, O. A. M. Abdelraouf, N.S.S. Ang, Z. Dong, J. K.W. Yang, H. Liu
Institute of Materials Research and Engineering (IMRE), Agency for Science, Technology and Research (A*STAR), 2 Fusionopolis Way, Innovis, #08-03, Singapore 138634
E-mail: h-liu@imre.a-star.edu.sg

H.T. Wang, H. Wang, Q. Ruan, W. Zhang, J. K.W. Yang
Singapore University of Technology and Design, Engineering Product Development, 8 Somapah Road, Singapore, 487372
E-mail: joel_yang@sutd.edu.sg

O. A. M. Abdelraouf
School of Physical and Mathematical Sciences, Nanyang Technological University, Singapore 637371, Singapore

Keywords: 3D structural colors, woodpile photonic crystals, high-order photonic cavity modes, two-photon lithography

It remains a challenge to directly print three-dimensional arbitrary shapes that exhibit structural colors at the micrometer scale. Woodpile photonic crystals (WPCs) fabricated via two-photon lithography (TPL) are promising as building blocks to produce 3D geometries that generate structural colors due to their ability to exhibit either omnidirectional or anisotropic photonic stopbands. However, existing approaches have focused on achieving structural colors when illuminating WPCs from the top, which necessitates print resolutions beyond the limit of commercial TPL and/or post-processing techniques. Here, we devised a new strategy to support high-order photonic cavity modes upon side-illumination on WPCs that surprisingly generate large reflectance peaks in the visible spectrum. Based on that, we demonstrate one-step printing of 3D photonic structural colors without requiring post-processing or subwavelength features. Vivid colors with reflectance peaks exhibiting a full width at half maximum of ~25 nm, a maximum reflectance of 50%, gamut of ~85% of sRGB, and large viewing angles, were achieved. In addition, we also demonstrated voxel-level manipulation and control of colors in arbitrary-shaped 3D objects constituted with WPCs as unit cells, which has great potential for applications in dynamic color displays, colorimetric sensing, anti-counterfeiting, and light-matter interaction platforms.

## 1. Introduction

In comparison to dyes or pigments, structural colors[1-5] exhibit outstanding attributes including high-brightness, fade-resistance, iridescence, and eco-friendliness.[6,7] These attributes are attractive for certain applications e.g. food dyes,[8] automobile paints,[9,10] pigment-free cosmetics,[11] optical display devices,[12-13] and anti-counterfeiting products.[14] With the development of micro- and nano-fabrication technologies, artificially engineered and miniaturized high-resolution structural colors have been employed effectively as fundamental elements for wearable optical displays and on-chip optical devices, such as user-interactive displays and micro-spectrometers.[15,16] Metallic nanostructures supporting plasmonic resonances[17-19] and dielectric nanoantennas with Mie scattering[20-23] are promising candidates for realizing such miniaturized optical components due to their bright and tunable structural colors. Additionally, subwavelength nanostructures made of magnetic materials[24], phase change materials[25], and polymers[26,27], have also been reported to achieve tunable 2D structural colors on the surface of various substrates. In contrast, realization of the full potential of structural colors in 3D provides greater freedom to shape, control, and display colors, beyond the constrains of 2D color printing.[28,29]

Woodpile photonic crystals (WPCs) are one of the most versatile platforms to realize 3D structural colors due to their rich and unique properties[30-33], particularly photonic band gaps (PBGs)[34], which can be simultaneously and independently engineered in three dimensions, enabling anisotropic control of structural colors. Unfortunately, despite the multitude of WPCs demonstrated in different materials, most of their PBGs occur at longer wavelengths beyond the visible region. The realization of visible frequency PBG requires subwavelength periodicity of WPCs, which is a challenge for conventional 2D nanofabrication techniques due to a combination of limited structural rigidity and process complexity.[28,33,35-39] Two-photon lithography (TPL) introduces design flexibility and feasibility in the fabrication of such WPCs enabling novel applications, such as 3D topological photonics (Weyl points),[40] energy dissipation and protective materials,[41] in addition to 3D structural colors. Previous studies on structural colors of WPCs have been focused on producing visible PBGs on the top or bottom surfaces, as light propagates along the stacking direction.[32,33,35] To achieve visible stop bands along this direction under top illumination, one has to reduce the lattice constants significantly below 500 nm (Figure S1 in Supporting Information) in all directions, which is beyond the lateral resolution limit of commercial TPL and the IP-Dip resin (~500 nm)[42]. Such high-resolution printing requires modified resists or complex optical setup (e.g., stimulated emission depletion direct laser writing, STED-DWL)[24,42,43], and multiple fabrication processes (alternating e-beam lithography and material deposition)[37]. Alternatively, one could apply post-processing techniques, e.g. heat shrinking (thermolysis) process to produce band stops and slow-light modes in the visible spectrum[32]. However, more investigations are needed to study the lateral response of WPCs, as light incident on the sides of the woodpile could potentially propagate along the constituent rods (i.e., along $\Gamma - K$ in the Brillouin zone, corresponding to side illumination) and support new mechanism to attain high performance structural colors.

Here, we theoretically and experimentally investigate the band structures of WPCs upon illumination from the side and reveal that specific photonic cavity modes can only be excited under side-illumination while they are not attainable from top-illumination. These modes are produced by the coupling between the cavity oscillation mode and high-order propagation modes along the horizontal rods. Interestingly, these photonic cavity modes can be tuned in the visible region, and simultaneously achieve excellent color purity, enhanced brightness with reflectance larger than 50%, as well as broad color gamut (larger than 85% of the sRGB space) upon variation of the lattice constants in the range of 0.9 µm-1.5 µm. These results validate our design strategy as an effective way to produce bright colors that can be printed in one step using TPL without the need for subwavelength lattice constants.

## 2. Results and discussion

### 2.1 High-order Photonic Cavity Propagation Modes

**Figure 1** illustrates the conceptual design and working principle of WPCs to generate colors upon the side-illumination. They consist of orthogonally stacked polymerized rods with an elliptical cross section due to the elongated voxel along the optical axis. Each unit cell is composed of four adjacent layers of rods stacked along $z$-axis with an in-plane pitch ($P_{xy}$) and an out-of-plane pitch ($P_z$). The rods of neighboring layers are orthogonal to each other, while those in every other layer are parallel to each other with an offset of $P_{xy}/2$, as shown in Figure 1a. The PBGs of WPCs can be calculated by solving the wave function equation within the Brillouin zone: $\frac{1}{\varepsilon(\boldsymbol{r})} \nabla \times \left( \frac{1}{\mu(\boldsymbol{r})} \nabla \times \boldsymbol{E} \right) - \frac{\omega^2}{c^2} \boldsymbol{E} = 0$, where $\varepsilon$ and $\mu$ denote the permittivity and permeability of the material at vector $\boldsymbol{r}$ within the PCs, respectively; $c$, $\omega$, and $\boldsymbol{E}$ denote the speed of light in vacuum, the angular frequency and the electric field, respectively[44]. Thus, the PBGs are mainly defined by keeping $\mu = 1$ and the periodically varying $\varepsilon(\boldsymbol{r})$, which is dependent on $P_{xy}$, $P_z$, and the refractive index contrast between the rods and surrounding environment. The side-illumination refers to the normal incidence on the side-surfaces of WPCs and the incident direction is denoted by the white arrow in Figure 1a.

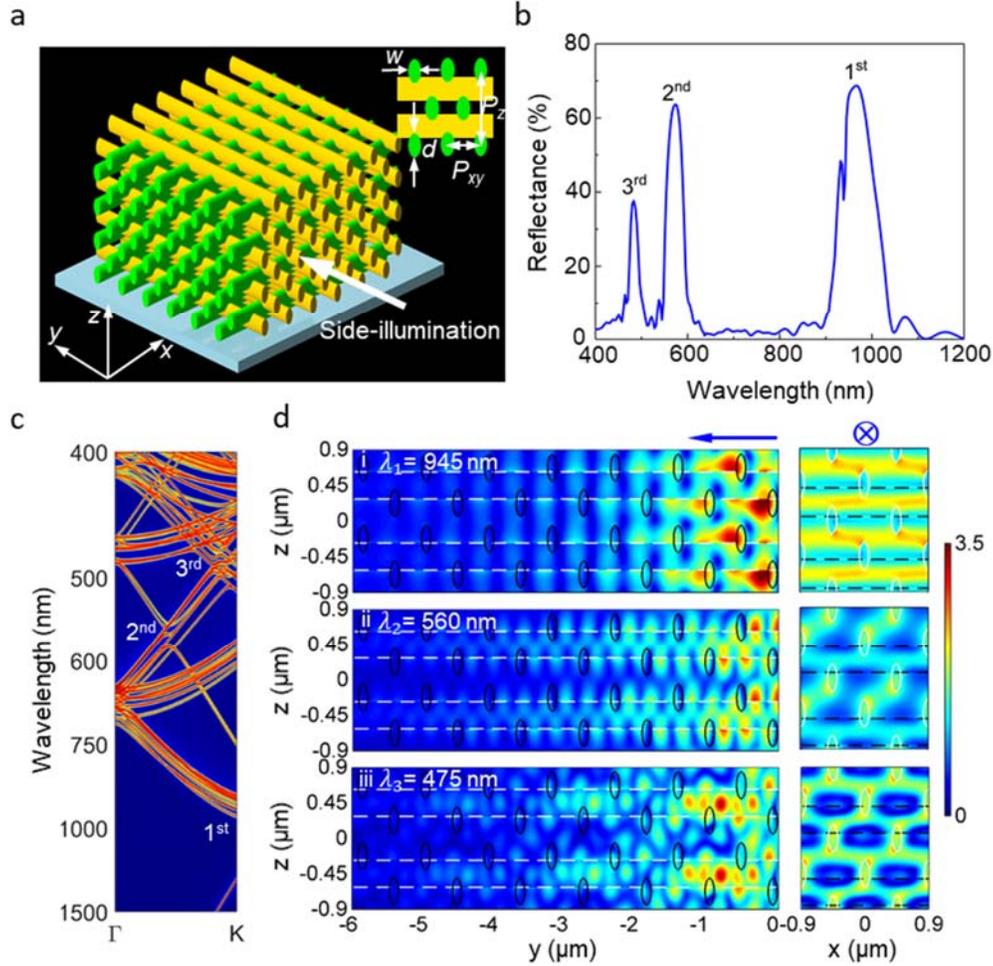

**Figure 1. Schematic of 3D visible photonic crystals.** (a) Structural design of WPCs. Side-illumination (white arrow) indicates the wave vector of the incident light parallel to the *y*-axis. Inset shows the side-view of a unit cell, consisting of rods with elliptical cross sections with major axis *d* and minor axis *w*. $P_{xy}$ and $P_z$ represents in-plane and out-of-plane pitch, respectively. (b) Calculated reflectance spectra of the WPCs with $P_{xy}$ = 0.9 µm, $P_z$ = 0.9 µm, *d* = 130 nm, *w* = 340 nm. Upon side-illumination, the 1st - 3rd reflectance peaks occur at the wavelengths of 945, 560, and 475 nm, respectively. Incident light is unpolarized. (c) Band structure of the WPCs along the *Γ-K* direction in the first Brillouin zone under side-illumination. The corresponding positions for the reflectance peaks are marked with ordinal numbers. (d) Calculated electric-field distributions along the *yz* and *xz* planes for the 1st (i), 2nd (ii), and 3rd mode (iii), respectively. The dashed lines and ellipses represent the rods along *x* or *y* axis that intersect the planes of plot. The blue arrow and the sign ⊗ indicate the propagation direction of light.

We calculated the reflectance spectra of the WPCs upon side- and top-illumination (the results for top-illumination is shown in Figure S2, Supporting Information). Figure 1b shows the simulated reflectance spectrum of WPCs with $P_{xy} = P_z = 0.9$ µm (As $P_z \neq \sqrt{2}P_{xy}$, the WPCs have a face-centered-tetragonal symmetry[45]) under side-illumination. A strong resonance occurs at ~945 nm, which is ascribed to the 1st stop band of the WPCs based on the calculated band structures (marked with the corresponding ordinal number in Figure 1c). Transmission spectrum (Figure S2, Supporting Information) and the near-field electrical-field distributions (Figure 1d-1i) were also calculated. As this peak occurs in the near-infrared region, it does not contribute to the structural colors of WPCs. In addition to the 1st stop band, another two intense reflection peaks are observed in the visible region, i.e., at 560 nm and 475 nm, respectively. These two resonances can be explained by the presence of slow light modes along $\Gamma$-$K$ direction in the first Brillouin zone, as marked by the ordinal numbers of 2nd and 3rd in Figure 1c. In the narrow wavelength near 560 nm and 475 nm, the group velocities of light at these slow modes are zero, weakening the coupling between incident light and WPCs due to the momentum mismatch, i.e., inducing narrow reflective peaks in the spectra. For simplicity, we termed these reflectance peaks from longer to shorter wavelength under side-illumination as the 1st, 2nd, and 3rd photonic cavity modes, respectively. Notably, the 2nd photonic mode (~560 nm) exhibits a narrow full width at half maximum (FWHM) of 40 nm and a high reflectance of 65%, indicating the high purity and brightness corresponding to yellow. Meanwhile, it overlaps with the 3rd photonic cavity mode (blue), leading to a green appearance. In contrast, as shown in Figure S2 (Supporting Information) under top-illumination, we observed only a small reflectance of less than 4% at 463 nm, which corresponds to the weak slow light modes along the $\Gamma$-$K$ direction. Such reflectance is too low to produce a visible color from the top surface of the WPCs. Based on the comparative analysis, it indicates that high-order photonic cavity modes can only be excited upon side-illumination, which is the key to produce high-quality reflective colors. However, such high-order modes are unattainable upon top-illumination, which inevitably leads to poor reflective color.

The electromagnetic field distributions within the WPCs were calculated using commercial finite-difference time-domain (FDTD) software to investigate the origin of the visible optical resonances (the 2nd and 3rd photonic cavity modes) upon the side-illumination. Figure 1d plots the electric-field distributions along the planes of $yz$ ($x$ = 0 µm) and $xz$ ($y$ = 0.9 µm) at the resonant wavelengths of those three photonic cavity modes. The light is normally incident on the $xz$ plane and propagates along the $y$-axis, as shown by the blue arrow in Figure 1a. Panel i in Figure 1d shows the electric-field map within the $yz$ plane of the 1$^{st}$ photonic cavity mode, indicating that the intensity of incident beam attenuates when the electric field propagates as the fundamental transverse electric mode through the cavities along the $y$ axis. After propagating a distance of 6 µm, the amplitude of electric field significantly drops to 5% of the incident beam, which indicates the absence of a complete PBG, verified by the transmission spectra (Figure S2, Supporting Information). The incomplete bandgap is due to the low refractive index ($n$ = 1.55) of the polymers, which is smaller than the required refractive index difference compared with the environment ($\Delta n$ > 1.9) for realizing complete bandgaps with WPCs[46]. Furthermore, referring to Figure S4 (Supporting Information), the propagation length is proportional to the reflectivity of the 1$^{st}$ photonic cavity mode, while it is inversely proportional to the transmittance. The electric-field distribution within the $xz$ plane shows that light is largely confined within the cavities instead of propagating along the horizontal rods as the minor axis ($w$) is only 130 nm, which is much smaller than the resonant wavelength of 945 nm.

Panel ii in Figure 1d is a plot of the electric-field distribution of the 2$^{nd}$ photonic mode at the resonant wavelength of 560 nm, which shows a high-order mode with an integer multiple of nodes/antinodes within each lattice. As the size ($d$ = 340 nm) of rods is comparable to the wavelength, it can be thought that light is initially guided to propagate along the $y$-axis rods (see electric-field distribution in the $yz$ plane). However, it also interacts with the rods along the $x$-axis that acts in a manner reminiscent of a distributed Bragg grating, resulting in a strong reflection thus forming the 2$^{nd}$ stop band. The electric-field distribution in the $xz$ plane clearly shows the coupling between the propagation mode (locally confined inside the rods) and the cavity mode. The effective propagation length of the 2$^{nd}$ photonic cavity mode is ~5 µm, which is relatively shorter than that of the 1st photonic cavity mode, and the electric-field intensity is decreased almost to zero after propagating such a distance. Thus, the reflectance peak intensity of 2nd mode does not increase until the length of WPCs along *y*-axis reaches 8 μm (Figure S3, Supporting Information). As the light is incident from the side of WPCs, the structure is similar to a distributed Bragg reflector with gratings patterned along the top and bottom surfaces of waveguides.

Panel iii in Figure 1d plots the corresponding electric-field distribution of the 3rd mode at ~475 nm, which is mainly attributed to the propagation mode within the *y*-axis rods under side-illumination, as the size (*d*) is almost comparable to the resonant wavelength. However, under top illumination (Figure S2, Supporting Information), a very weak peak at ~463 nm can be seen, and its electric field (Figure S4, Supporting Information) is mainly distributed on the surface of the rods within the cavity, defined as the space between neighbouring rods. The absence of the 2nd and 3rd photonic cavity modes under the top-illumination is resulted from the physical arrangement of the rods of WPCs along horizontal and longitudinal directions (i.e., within *xy* planes). The absence of through rods along the vertical direction explains the extremely weak or absence of the higher-order modes (e.g., the 2nd and 3rd modes) upon top-illumination. In comparison, the strong high-order modes are excited under side-illumination to produce high reflectance, resulting in the observed structural colors on the four sides of WPCs with such large periodicities.

Next, we investigated the effect of structural parameters and incidence angle on the high-order photonic cavity modes and the results of theoretical calculations are shown in **Figure 2**. Figure 2a shows the reflectance spectra of high-order photonic cavity modes as $P_{xy}$ was varied from 0.8 μm to 1.3 μm keeping $P_z$ constant at 0.9 μm. The resonant peak of the 2nd mode red shifts from 527 nm to 730 nm as $P_{xy}$ increases, providing evidence for the Bragg interference nature of these peaks. The reflectance peaks remain above 60% till $P_{xy}$ reaches 1.2 μm and it drops to 50% when $P_{xy}$ rises to 1.3 μm. Simultaneously, the 3rd- and 4th-order photonic cavity modes gradually red shift in the visible region and both of their maximal reflectance is larger than 50%. Figure 2b plots the calculated results upon increasing $P_z$ from 0.9 µm to 1.4 µm with $P_{xy}$ fixed at 0.9 µm. Notably, the redshift of the peak wavelength of the 2nd and 3rd photonic cavity modes is around 50 nm and 15 nm respectively, which is much less than that modulated by $P_{xy}$, as shown in Figure 2a. In terms of amplitude of the resonant intensities, the results for the 2nd mode in Figures 2a and 2b are comparable, while those of 3rd and 4th modes are relatively lower in Figure 2b. These results can be explained by the distributed Bragg reflector nature of woodpile structures. When $P_{xy}$ increases, the space within each cavity along the propagation direction becomes larger so that higher-order propagation modes of photonic cavities are supported. In comparison, the increase of $P_z$ has little effect on the resonant wavelength of the higher-order cavity modes as it is perpendicular to the incidence beam. It shows that high-order photonic cavity modes is more sensitive to the variation of $P_{xy}$ than $P_z$.

Figure 2c presents the dependence of the frequency and intensity of the 2nd and 3rd photonic cavity modes upon varying $d$ and $w$, respectively. Here, $P_{xy}$ and $P_z$ are both fixed as 0.9 µm. As $d$ increases from 280 nm to 480 nm, the frequency and intensity of the 2nd and 3rd photonic cavity modes are slightly changed due to the fact that perpendicular rods are intersecting with each other at the cross points to enhance the Bragg condition. Conversely, in the case of increasing $w$ from 100 nm to 250 nm, the amplitude for both modes gradually drops and eventually vanishes. As the minor axis ($w$) increases, the redshift of the modes indicate that the effective refractive index of the system has increased as the rods expand into surrounding air. Upon side-illumination, WPCs work similar to a distributed Bragg reflector with gratings atop waveguides. The disappearance of the reflectance peaks occurs when the rods tend to function as an effective medium with a network of narrow holes rather than waveguides. Therefore, keeping the minor axis ($w$) of the rods as small as possible is a key factor to determine the brightness (reflectance) of the structural colors upon side-illumination.

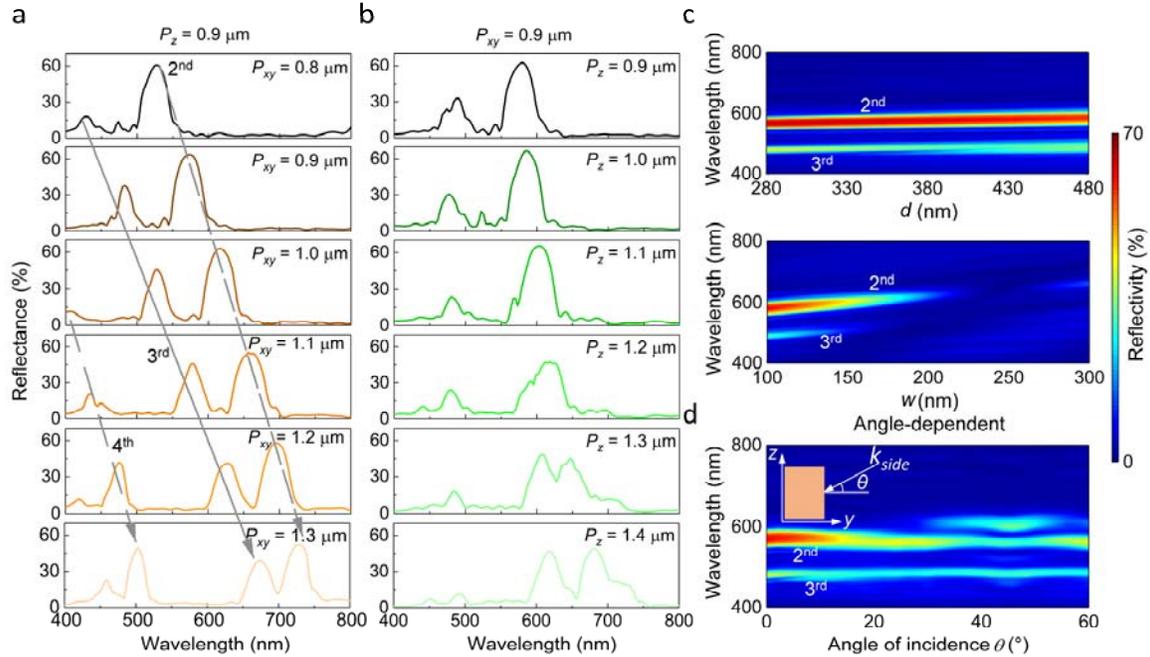

**Figure 2. Influence of structural parameters on photonic cavity propagation modes.** (a, b) Simulated reflective spectra of WPCs with $P_z$ = 0.9 µm, and $P_{xy}$ = 0.9 µm, respectively. $d$ and $w$ were fixed as 340nm and 130 nm, respectively in (a). Considering the experimentally feasible feature sizes, $d$ and $w$ in (b) were fixed as 400 nm and 130 nm respectively. (c) Simulated reflectance mapping of the 2$^{nd}$ and 3$^{rd}$ photonic cavity modes as a function of $d$ ($w$ = 140 nm) and $w$ ($d$ = 340 nm), with $P_{xy}$ and $P_z$ both fixed at 0.9 µm. (d) Angle-resolved reflectance of the 2$^{nd}$ and 3$^{rd}$ photonic cavity modes upon-side illumination. $k_{side}$ represents the direction of wave vector of the incident light, indicated by the angle ($\theta$) with respect to $y$-axis.

Moreover, we also investigated the angle-dependence of the 2$^{nd}$ and 3$^{rd}$ photonic cavity modes and the results are shown in Figure 2d and Figure S5 (Supporting Information), respectively. With the angle of incidence ($\theta$) rising from 0° to 30°, the resonant wavelength of the 2$^{nd}$ mode exhibits a small blue shift of ~10 nm and its intensity drops less than 15%. When $\theta$ continuously increases from 30° to 45°, the reflectance spectrum gradually splits into two peaks, which means the variation of the structural colors of WPCs. The 3$^{rd}$ mode is insensitive to the variation of $\theta$, which can be attributed to its resonant modes locally confined within the rods along the light propagation direction. The angle-resolved reflectance spectra in Figure 2d indicates that the structural colors of WPCs upon side-illumination are relatively insensitive to the incident angle within ±30°, manifesting uniform structural colors over a large viewing angle.

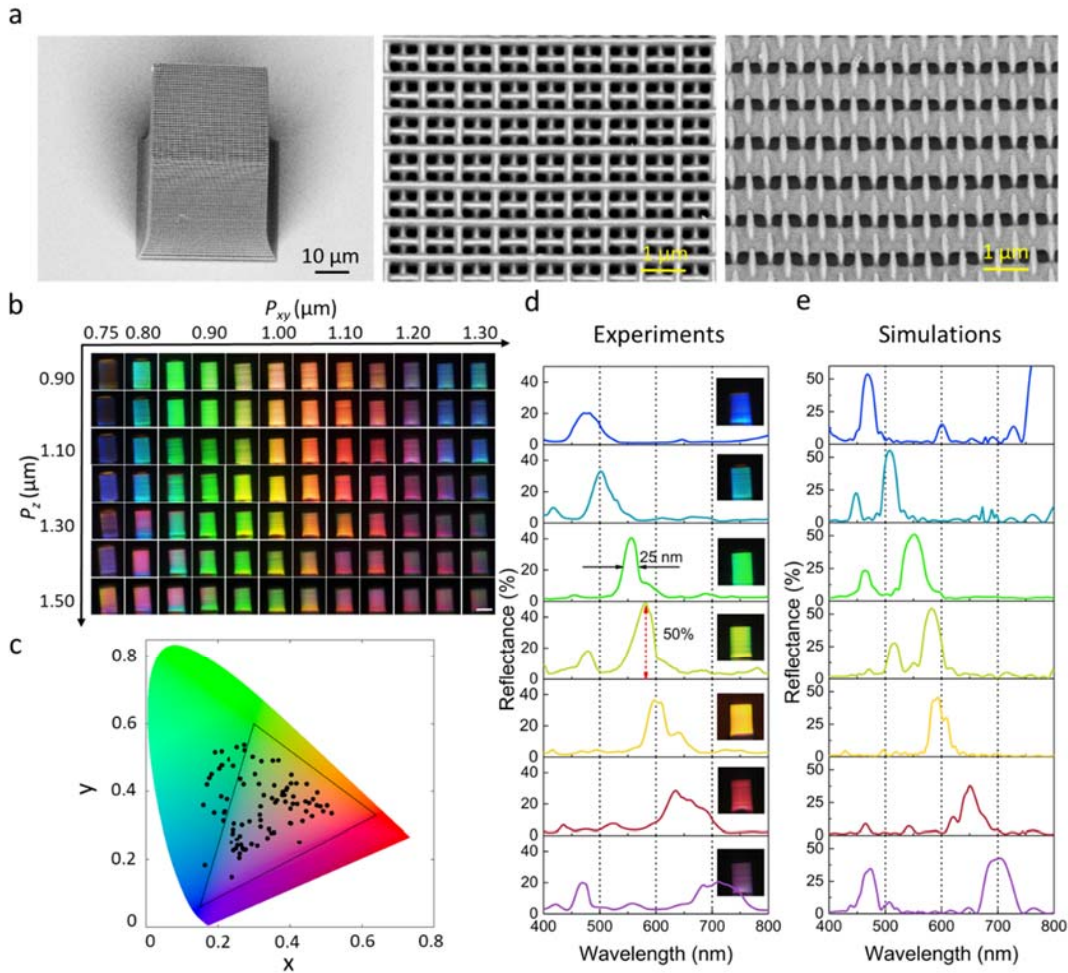

**Figure 3. Side-view WPC structural colors.** (a) SEM images of a printed WPC (left panel) with a size of 40 μm (width) × 40 μm (length) × 60 μm (height). Zoom-in SEM images of the top (middle panel) and the side (right panel). (b) Color palette of the side-views of the printed WPCs with $P_{xy}$ varying from 0.75 to 1.3 μm and $P_z$ from 0.9 to 1.5 μm. Scale bar: 40 μm. (c) Measured color coordinates of the fabricated WPCs plotted as dots on CIE 1931 color space chromaticity diagram. (d) Measured and (e) simulated reflectance spectra of the blue, cyan, green, green-yellow, yellow, red, and purple colors. Insets in (d) are the corresponding optical images.

**2.2 Photonic Structural Colors.** To validate the design strategy elaborated above, we fabricated a multitude of WPCs with $P_{xy}$ varying from 0.75 μm to 1.3 μm and $P_z$ from 0.9 μm to 1.4 μm. A commercial TPL system (Photonic Professional GT2, Nanoscribe, GmbH) together with its low refractive index resin $n$ = 1.55 (IP-Dip, GmbH)[47] was used. The fabrication process is detailed in Methods. **Figure 3**a shows scanning electron microscope (SEM) images of a fabricated sample with $P_{xy}$ = 0.9 μm, $P_z$ = 1.0 μm, respectively. The top-view (middle panel) and side-view (right panel) clearly show the well-resolved unit cells. The polymerized rods exhibit elliptical cross section with the measured depth (*d*) of 380 nm and the width (*w*) of 130 nm, respectively. As aforementioned, when the width of the rods was within the range of 100 nm to 200 nm (Figure 2c), the high-order photonic cavity mode of WPCs enables high reflectance to acquire high brightness structural colors. Figure 3b shows the captured optical micrographs of WPCs with different pitches upon side-illumination, where the samples were imaged under an optical microscope with a 10× objective lens and numerical aperture (NA) of 0.15, with its side facing upwards with respect to the objective lens to ensure normal incident. A wide range of colors including purple, blue, green, cyan, yellow, orange, pink and red was achieved. To quantitatively characterize the gamut of these structural colors, we measured their spectra and mapped their coordinates in the CIE 1931 color space chromaticity diagram, as shown in Figure 3c. It shows that the color coordinates occupy more than 85% of the sRGB color space, with some green and blue colors lying beyond the sRGB triangle. Larger color gamut could be achieved upon optimizing the lattice constants, structural parameters, or even varying structural periodicities. In addition, the brightness and saturation of the colors increases as $P_z$ decreases, which is in agreement with the calculations in Figure 2b.

To quantify the purity and brightness of the printed structural colors, we measured the reflectance spectra of blue, cyan, green, yellow-green, yellow, red, and purple colors and the results are plotted in Figure 3d. The 2$^{nd}$ mode gradually red shifts from 460 nm to 730 nm upon the increase of pitch size. Due to the high-order mode, we achieved a FWHM of ~25 nm which is smaller than that of Si metasurfaces (FWHM of ~35-40 nm)[48-50], indicating a high purity (hue) of the reflective colors. Moreover, the maximum reflectance is ~50%, resulting in the high brightness of colors. Upon increasing $P_{xy}$, more high-order (3$^{rd}$ and 4$^{th}$ orders) modes start to form in the visible region. When both of them appear simultaneously, it leads to a mixture of colors, e.g., purple (as shown in the bottom Panel and its inset). Figure 3e plots the simulated results which are in good agreement with the measurement. Overall, the achieved structural colors upon side illumination demonstrate the features of broad color gamut, high purity and brightness, which has great potential for dynamic color display.

## 2.3 Miniaturized Photonic Merlions

We have demonstrated a novel strategy to excite lateral high-order photonic cavity modes in a typical WPC, which is employed as the elementary unit to directly print high-quality structural colors in 3D manner. To further explore its applicability towards printing 3D free-form structural colors, we have adopted the strategy to print monochromatic and multicolor 3D arbitrary-shaped structures in one step. The details of design and fabrication processes are described in the methods.

A miniaturized Merlion, an iconic mascot of Singapore, was employed as an example of monochromatic 3D structures. **Figure 4**a shows the SEM images of a printed Merlion viewing from different angles. The dimensions along *x*, *y*, and *z* directions are 140 × 95 × 247 μm$^3$, and the periodicities of $P_{xy}$ and $P_z$ are 1.0 μm and 0.9 μm, respectively, confirmed by the zoom-in SEM image. To achieve different colors, we fabricated a series of Merlions with different geometrical parameters, showing purple, blue, cyan, green, yellow, and pink colors, as shown in Figure 4b. These structural colors exhibit significant spatial uniformity, high purity, and high reflective brightness. A fabricated green Merlion ($P_{xy}$ = 0.85 μm, $P_z$ = 1.0 μm) was rotated along its *z*-axis under side-illumination. The captured optical micrographs (Figure S6, Supporting Information) show uniform green colors across the majority of the surfaces. Inconsistence was only observed at some small parts including the base showing blue colors and some of the hair part at right and back sides showing red colors, which is related to the slight drift of the laser power during fabrication process.

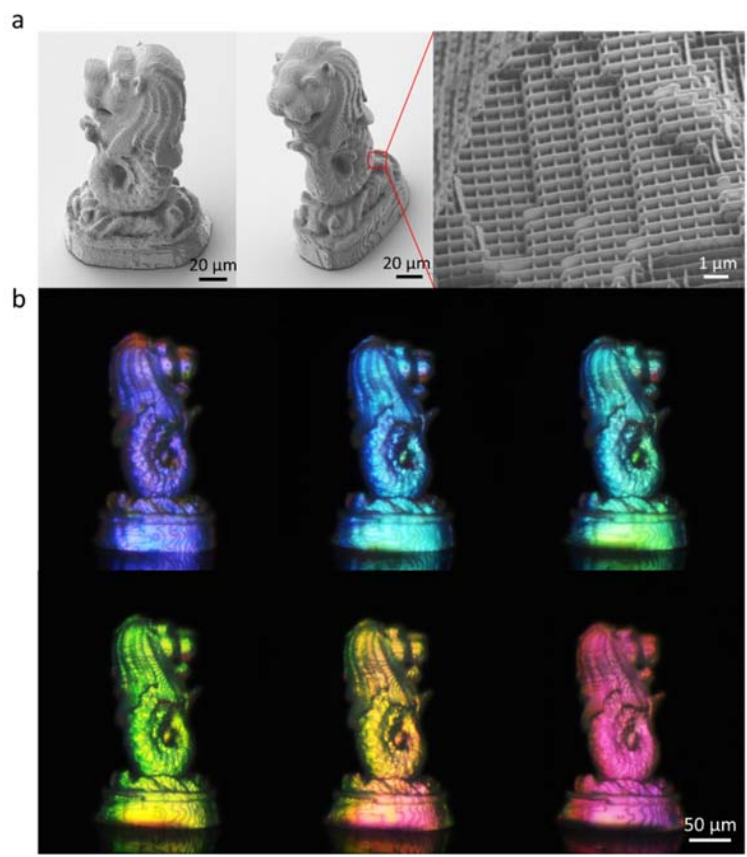

**Figure 4. 3D Photonic Merlions with uniform monochromatic structural colors.** (a) SEM images of a fabricated Merlion under different viewing angles (left and middle panels). A zoom-in SEM image (right panel). (b) Miniaturized 3D Merlions with monochromatic structural colors printed by TPL. The geometrical parameters for different structural colors are as the follows: purple ($P_{xy}$ = 0.75 µm, $P_z$ = 0.8 µm), blue ($P_{xy}$ = 0.85 µm, $P_z$ = 1.1 µm), cyan ($P_{xy}$ = 0.8 µm, $P_z$ = 0.9 µm), green ($P_{xy}$ = 0.95 µm, $P_z$ = 1.3 µm), yellow ($P_{xy}$ = 1.0 µm, $P_z$ = 1.2 µm), and pink ($P_{xy}$ = 0.8 µm, $P_z$ = 1.4 µm). Scale bars: 50 µm.

In addition, we investigated the ability of precise color tuning at the voxel level with this approach. Figure 5a shows a miniaturized Merlion printed with gradually varying structural colors under reflection mode —purple head, red hairs, eyebrows, and nose, blue neck and body, and gray base. The complex color tuning was realized by the gradual change one of two periodicities (here, we varied $P_{xy}$ while fixing $P_z$) of the corresponding voxels in WPCs.

In addition to monochromatic Merlion and gradually varying its color, we also demonstrated the abrupt tuning of colors (i.e., simultaneously varying $P_{xy}$ and $P_z$ between two neighboring voxels), which is a prerequisite to realize arbitrary-shaped 3D objects with colors at will. A miniaturized "Rubik's cube" with abruptly changed structural colors over its cube faces was printed, as shown in **Figure 5**. The top panel in Figure 5b shows its color image, taken from its diagonal direction under the incident light perpendicular to the common edge between faces 1 and 2. The bottom panel in Figure 5b shows the front-view micrographs of these two faces under the normal incidence. We can clearly see the colors abruptly changed between neighboring small cubes that are due to the simultaneously changed $P_{xy}$ and $P_z$ (e.g., $P_{xy}$ = 1.1 µm, $P_z$ = 0.9 µm for cube A in Face 1, and $P_{xy}$ = 1.0 µm, $P_z$ = 1.2 µm for cube B). The difference of colors as shown in top and bottom panels was caused by different incident conditions. Although the cube colors are iridescent under large incident angle, we still can accurately control the structural colors of each cube and abruptly change the colors of neighboring cubes.

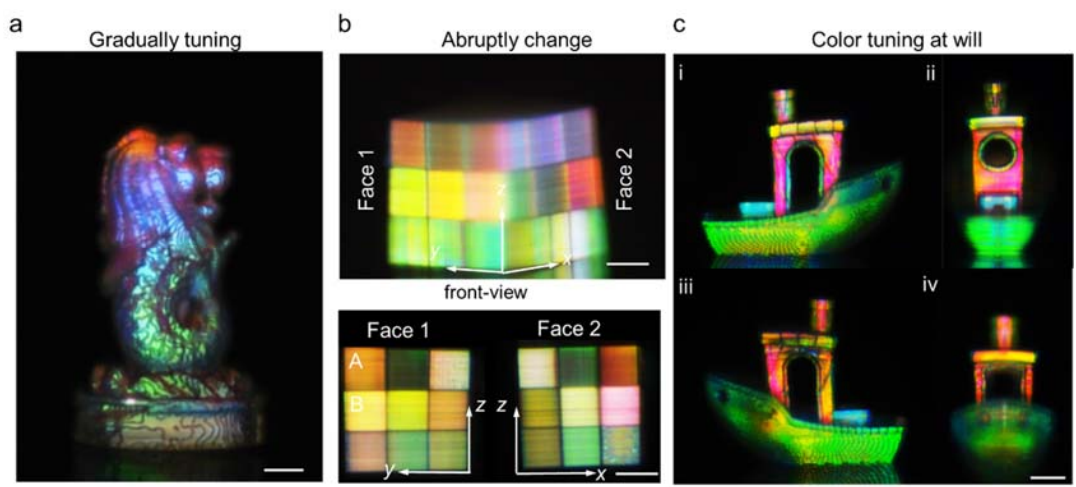

**Figure 5. Voxel level complex color tuning for 3D arbitrary-shaped structures.** (a) 3D photonic Merlion with gradually tuning colors. The structural colors from Merlion head to its tail are gradually changed from red ($P_z$ = 1.1 µm $P_{xy}$ =1.1 µm) to purple ($P_z$ = 1.1 µm $P_{xy}$ =1.2 µm), blue ($P_z$ = 1.1 µm $P_{xy}$ =1.3 µm) and then green colors ($P_z$ = 1.1 µm $P_{xy}$ =0.9 µm). Scale bar: 50 µm. (b) Photonic "Rubik's cube" with abruptly varying colors. The top side-view image was captured under oblique incidence, i.e., rotating the cube 45° along its *z*-axis with the incident light perpendicular to the edge between Faces 1 and 2. The front-view optical micrographs were taken with the incident light perpendicular to Faces 1 or 2. Scale bars: 30 µm. (c) The photonic Benchy with desired structural colors. (i-iv) Different sides of the printed Benchy. Scale bars: 30 µm.

In addition, the 3D Benchy model was also printed to demonstrate arbitrary-shaped 3D structures with pre-defined colors. 3D Benchy model has been commonly used as a benchmark for 3D printing performance, as it includes complex shapes such as curved surfaces, overhanging structures, holes, hollow objects, and etc. We assigned different structural colors to different components in this model. Figure 5c(i) shows the side-view optical micrograph image of a printed Benchy sample showing green hull, yellow roof, purple-red bridge and chimney, and blue arc and box on the deck. Even though the hull is a curved surface, we still observe uniform colors across most of the surface, thanks to the merit of non-sensitive angle-dependence of the high-order photonic cavity modes. In addition, we capture its structural colors from different sides of this printed Benchy, as shown in Figure 5c(i - iv). To the best of our knowledge, this is the first and smallest Benchy ($178 \times 92 \times 143$ $\mu m^3$) with different structural colors to date. We have demonstrated the ability of printing desirable colors on 3D complex structures by employing WPCs as unit cell and precisely tailoring its feature size via TPL, which can be extended to generate 3D free-form structural colors at will.

## 3. Conclusion

The generation of structural color in photonic crystals does not necessitate complete bandgaps or bandstops. Our strategy leverages on high-order photonic cavity modes in WPCs upon side-illumination, which effectively produces high performance structural colors without the requirements for high refractive index materials, sub-wavelength pitch, or post processing. The experimentally achieved structural colors exhibit large color gamut, high purity and reflectance, as well as good uniformity under a wide viewing angle, which outperform those results achieved under top illumination. Furthermore, by employing WPCs as unit cells, we have demonstrated the precise printing and tuning of colors at the voxel level on 3D arbitrary-shaped structures, leading to applications of coloration-based sensors, color displays, directional light emission devices, as well as anti-counterfeiting.

## 4. Experimental Section/Methods

*Fabrication of Woodpile Photonic Crystals and 3D Free-form Structures*. The woodpile photonic crystals and 3D structural colors were fabricated with a commercial two-photon lithography setup, Photonic Professional GT2 (Nanoscribe, GmbH). The photoresist was IP-dip resist (Nanoscribe, GmbH), which was firstly drop-casted onto a fused silica glass substrate (25 mm × 25 mm × 0.7 mm) before printing. The design and fabrication processes of 3D structural colors are described as the follows. Firstly, a 3D model was sliced layer by layer to generate the coordinates voxels, required for controlling the trajectories of the ultra-compacted focal spot of the femtosecond laser beam (wavelength of 780 nm, pulse of 100 fs, and repetition rate of 80 MHz). Secondly, structural color and its corresponding geometrical parameters were assigned to each voxel based on the measured color palettes in Fig. 3a. Thirdly, the generated geometrical parameters ($P_{xy}$, $P_z$, $d$, and $w$), together with laser writing speeds and powers, were programmed to control the scan speed and spot energy of the laser beam for each voxel, forming 3D profiles with the desired photonic colors. Finally, the exposed samples were sequentially developed in propylene glycol methyl ether acetate (Sigma-Aldrich) and isopropyl alcohol to rinse off the undeveloped resins, achieving the desired free-form objects with structural colors.

*Theoretical Simulations.* The simulated spectra, photonic band structures and near-field electric-field distributions were calculated with FDTD solution (Lumerical Int.). For side-illumination, periodic boundaries were applied along $x$ and $z$ directions, while perfectly matched layer (PML) boundary was used for the direction of $y$. Under top-illumination, $x$ and $y$ directions were periodic boundaries, and z direction was PML boundary. Unpolarized plane wave was employed as the light source, and the meshing parameters were set as 5 nm$^3$.

*Optical and SEM Characterization.* All the experimental spectra and the optical micrographs of the WPCs and 3D structures were measured with a Nikon microscope (LV100ND) equipped with a long-distance objective lens (10×, NA=1.5). The sample was fixed on an adjustable flip platform and the measured face of the WPCs or 3D structures was facing towards the objective lens to make sure that the light was perpendicularly incident to the measured surface. The spectra were measured with a micro-spectrometer (CRAIC) attached to the Nikon microscope.

The SEM images were collected with a scanning electric microscope (JSM-7600F, JEOL) with an acceleration voltage of 5 kV.

**Supporting Information**

Supporting Information is available from the Wiley Online Library or from the author.

**Acknowledgements**

This research is partially supported by IMRE fund and A*STAR AME programmatic grant (grant no. A18A7b0058), IMRE project (SC25/18-8R1804-PRJ8). Author contributions: H.L.L. conceived the idea, fabricated the photonic crystals and 3D structures, characterized the results, and simulated the reflectance spectra. H.T.W., H.W., Q.F.R., and W.Z. did the band structure simulations. J.D. and N.S.S.A. did the SEM characterization. H.L.L., J.K.W.Y. and H.L. drafted and finalized the manuscript. All the authors discussed the results and commented on the manuscript. H.L. supervised the project.

Received: ((will be filled in by the editorial staff))
Revised: ((will be filled in by the editorial staff))
Published online: ((will be filled in by the editorial staff))

ToC figure

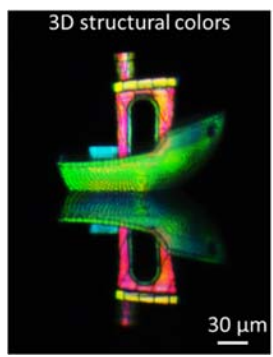

Supporting Information

**High-order Photonic Cavity Modes Enabled 3D Structural Colors**

*Hailong Liu, Hongtao Wang, Hao Wang, Jie Deng, Qifeng Ruan, Wang Zhang, Omar A. M. Abdelraouf, Noman Soon Seng Ang, Zhaogang Dong, Joel KW Yang\*, Hong Liu\**

H.L. Liu, J. Deng, O. A. M. Abdelraouf, N. S. S. Ang, Z. Dong, J. K.W. Yang, H. Liu
Institute of Materials Research and Engineering (IMRE), Agency for Science, Technology and Research (A*STAR), 2 Fusionopolis Way, Innovis, #08-03, Singapore 138634
E-mail: h-liu@imre.a-star.edu.sg

H.T. Wang, H. Wang, Q. Ruan, W. Zhang, J. K.W. Yang
Singapore University of Technology and Design, Engineering Product Development, 8 Somapah Road, Singapore, 487372
E-mail: joel_yang@sutd.edu.sg

O. A. M. Abdelraouf
School of Physical and Mathematical Sciences, Nanyang Technological University, Singapore 637371, Singapore

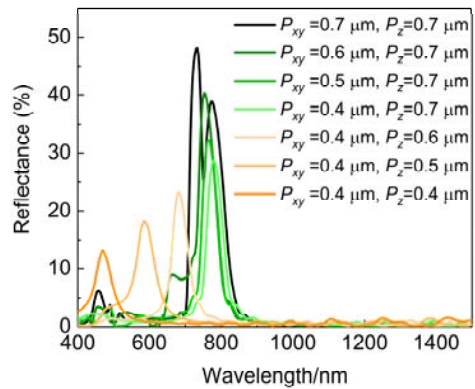

**Figure S1.** Simulated reflectance spectra of woodpile photonic crystals with visible bandgaps under the normal illumination on the top. (a) Reflectance spectra with varying $P_{xy}$ and $P_z$. Simulation results demonstrate the bandgap mainly depends on $P_z$, and the bandgap blue shifts to the visible region as $P_z$ is smaller than 0.6 μm. As the $P_z$ is 0.6 μm, the distance between two neighboring layers is $P_z/4 = 125$ nm, which is out of the range of the fabrication resolution limitation of TPL along the $z$-direction.

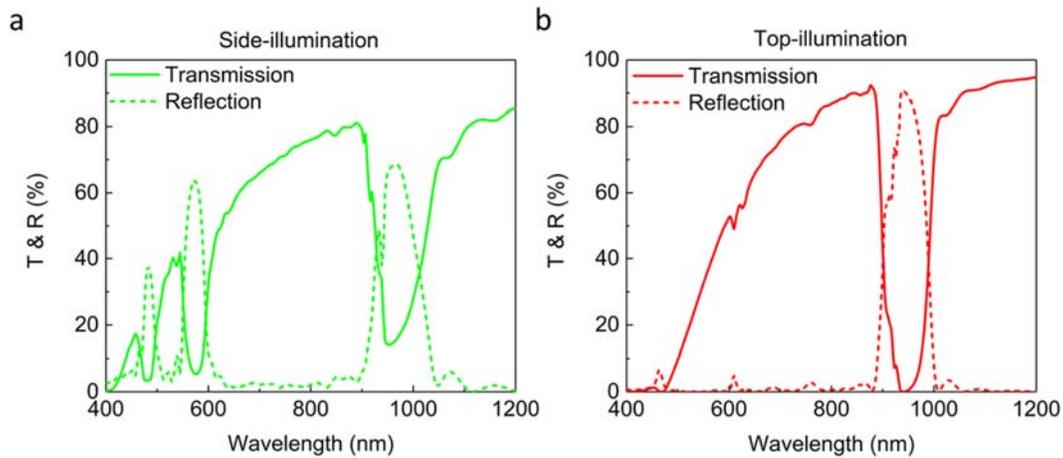

**Figure S2.** Simulated transmission and reflectance spectra of WPC under (a) side- and (b) top-illumination. The geometrical parameters of the WPC are detailed as follows: with $P_{xy} = 0.9$ μm, $P_z = 0.9$ μm, $w = 130$ nm, $d = 340$ nm.

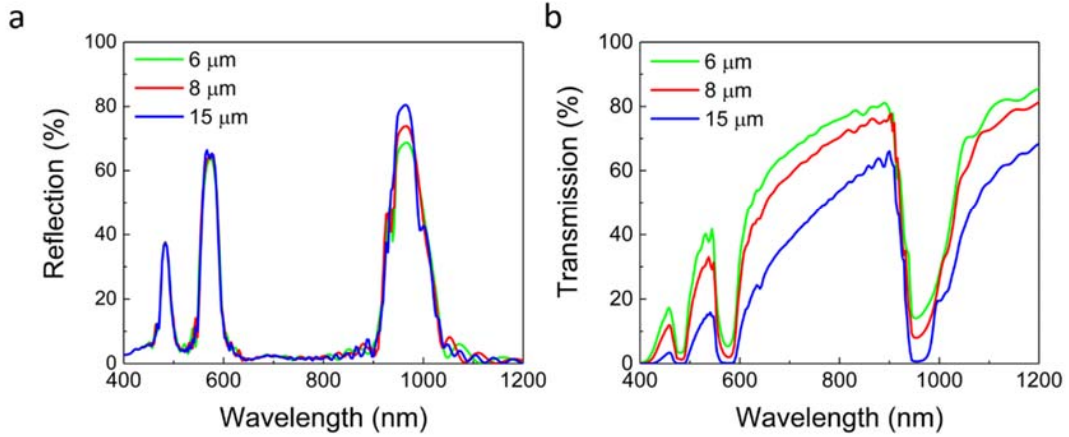

**Figure S3.** Influence of propagation length on stop bands of woodpile photonic crystals. Reflectance (a) and transmission (b) spectra of woodpile photonic crystals with different optical path under side-illumination. The thickness along $k_{side}$ (optical path) direction is 6, 8, and 15 µm, respectively. The geometrical parameters is $P_{xy}$ = 0.9 µm, $P_z$ = 0.9 µm, $w$ = 130 nm, $d$ = 340nm.

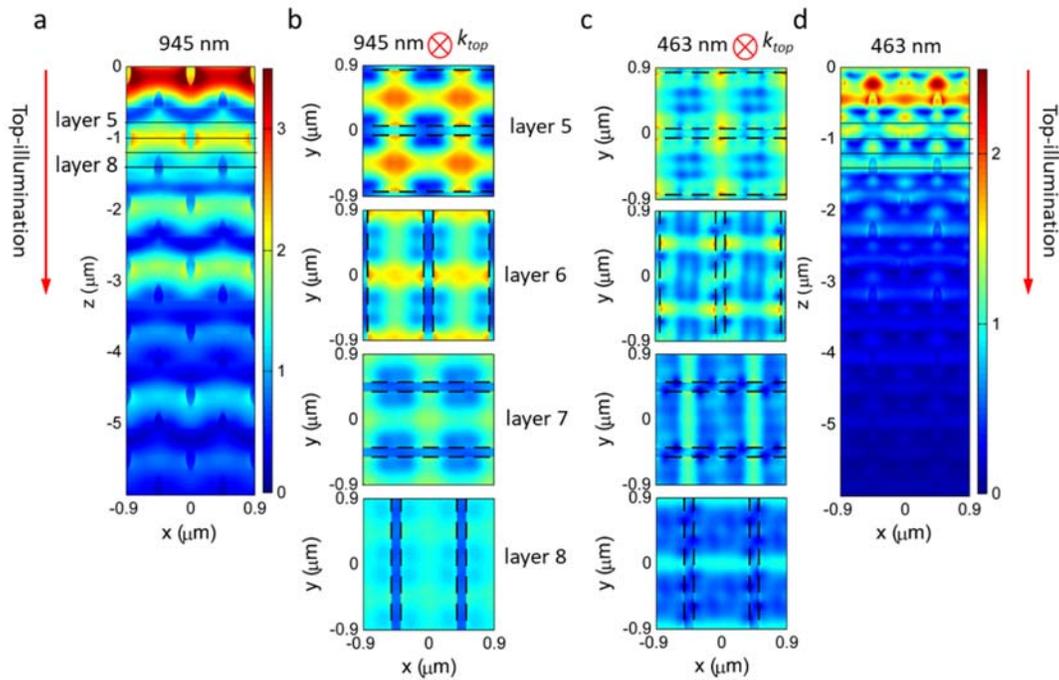

**Figure S4.** Near-field electric-field distributions of woodpile photonic crystals under top-illumination. The geometric parameters are $P_{xy}$ = 0.9 µm, $P_z$ = 0.9 µm, $w$ = 130 nm, $d$ =340 nm. (a) Electric-field distributions within $xz$ plane (a) and $xy$ planes (b) under the incident wavelength ($\lambda$) of 945 nm. The incident wavelength for (c, d) is 463 nm.

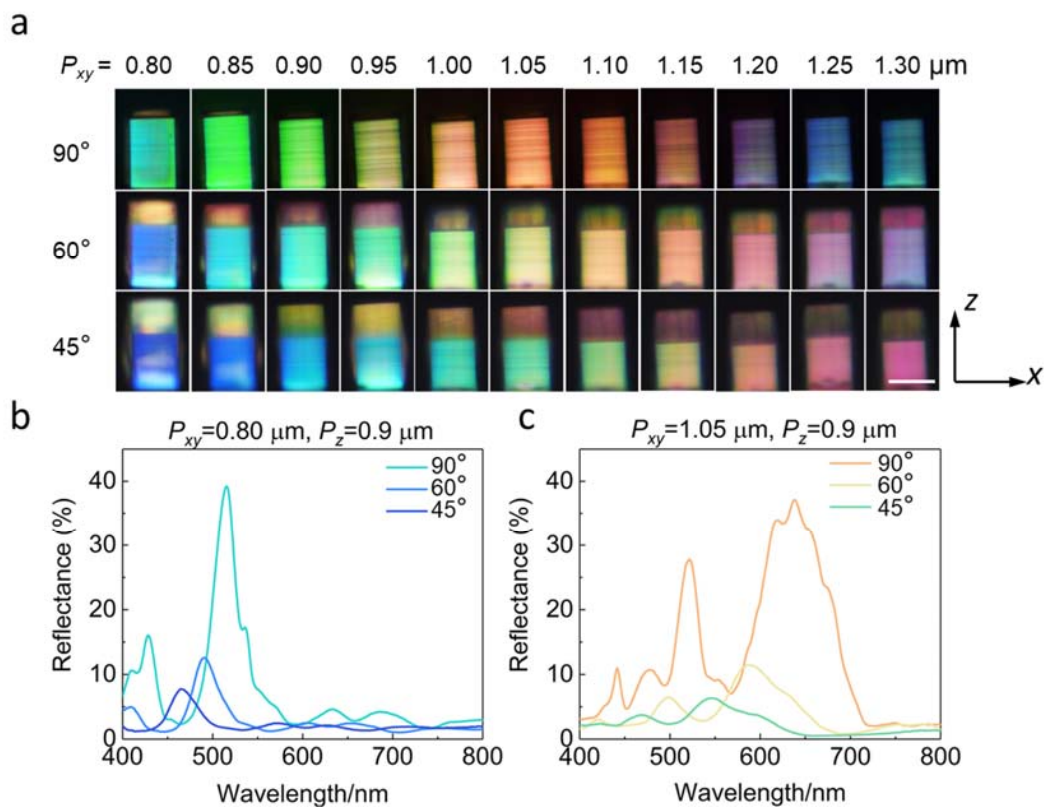

**Figure S5.** Angle-dependent side colors of woodpile PCs. (a) Angle-dependent side color patches of woodpile PCs with $P_{xy}$ varying from 0.8 µm to 1.3 µm and $P_z$ fixed as a constant of 0.9 µm. Measured reflectance spectra with $P_{xy}$ = 0.8 µm and $P_z$ = 0.9 µm (b) and $P_{xy}$ = 1.05 µm, $P_z$ = 0.9 µm (c) under different viewing angles. Here, the light is normally incident to the *xz* plane, and the woodpile is clockwise rotating along the *x*-axis at 45°, 60°, and 90°. Scale bar: 40 µm.

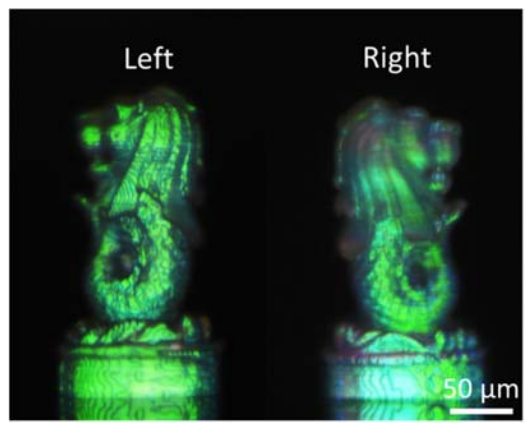

**Figure S6.** A 3D printed miniaturized green Merlion of side-views from different directions, showing extremely spatial uniformity of the structural colors. The geometrical parameters of the green 3D Merlion are: $P_{xy}$ = 0.85 µm and $P_z$ = 1.0 µm.

4